\documentclass{aastex62}

\usepackage{multirow}
\received{XXXX}
\revised{XXXX}
\accepted{XXXX}
\submitjournal{AAS}
\shorttitle{Detection of a White Dwarf companion to a Blue Straggler in M67}
\shortauthors{Sindhu et al.}

\begin{document}

\title{\textit{UVIT} Open Cluster Study. I. Detection of a White Dwarf Companion to a Blue Straggler in M67: Evidence of Formation through Mass Transfer}

\author[0000-0001-6006-1727]{N. Sindhu}
\affiliation{Department of Physics, School of Advanced Science, VIT, Vellore -632014, India}  
\affiliation{Indian Institute of Astrophysics, Koramangala II Block, Bangalore-560034, India}
\email{sindhu.n@iiap.res.in}

\author{Annapurni Subramaniam}
\affiliation{Indian Institute of Astrophysics, Koramangala II Block, Bangalore-560034, India}
%\email{purni@iiap.res.in}

\author[0000-0002-8672-3300]{Vikrant V. Jadhav}
\affiliation{Indian Institute of Astrophysics, Koramangala II Block, Bangalore-560034, India}
\affiliation{Joint Astronomy Program and Physics Department, Indian Institute of Science, Bangalore - 560012, India}

\author[0000-0002-3680-2684]{Sourav Chatterjee}
\affiliation{DAA, TIFR, Homi Bhabha Road, Navy Nagar, Colaba, Mumbai 400005, India}

\author{Aaron M. Geller}
\affiliation{Center for Interdisciplinary Exploration and Research in Astrophysics (CIERA) and Department of Physics \& Astronomy, Northwestern University, 2145 Sheridan Rd., Evanston, IL 60201, USA}
\affiliation{Adler Planetarium, Dept. of Astronomy, 1300 S. Lake Shore Drive, Chicago, IL 60605, USA}

\author{Christian Knigge}
\affiliation{School of Physics and Astronomy, University of Southampton, Highfield, Southampton SO17 1BJ, UK}

\author{Nathan Leigh}
\affiliation{Departamento de Astronom\'ia, Facultad de Ciencias F\'isicas y Matem\'aticas, Universidad de Concepci\'on, Concepci\'on, Chile}
\affiliation{Department of Astrophysics, American Museum of Natural History, New York, NY 10024, USA}

\author[0000-0003-0350-7061]{Thomas H.~Puzia}
\affiliation{Institute of Astrophysics, Pontificia Universidad Cat\'olica de Chile, Av.~Vicu\~na Mackenna 4860, 7820436 Macul, Santiago, Chile}

\author{Michael Shara}
\affiliation{Department of Astrophysics, American Museum of Natural History, New York, NY 10024, USA}

\author{Mirko Simunovic}
\affiliation{Gemini Observatory, 670 N Aohoku Pl, Hilo, Hawaii 96720}
\affiliation{Subaru Telescope, 650 N Aohoku Pl, Hilo, HI 96720, USA}

\begin{abstract}

The old open cluster M67, populated with blue straggler stars (BSSs), is a well known test bed to study the BSS formation pathways. Here, we report the first direct detection of a white dwarf (WD) companion to a BSS in M67, using far-UV images from the \textit{Ultra Violet Imaging telescope} ({\it UVIT}) on {\it ASTROSAT}. Near-simultaneous observations in three far-UV bands combined with {\it GALEX}, {\it IUE}, ground and space based photometric data covering 0.14 -11.5 $\mu$m range for WOCS1007 were found to require a binary fit to its spectral energy distribution (SED), consisting of a BSS and a hot companion. On the other hand, a single spectral fit was found to be satisfactory for the SEDs of two other BSSs, WOCS1006 and WOCS2011, with the latter showing a deficient far-UV flux. The hot companion of WOCS1007 is found to have a T$_{eff}$ $\sim$ 13250-13750K and a radius of 0.09$\pm$0.01 R$_\odot$. A comparison with WD models suggests it to be a low mass WD ($\sim$ 0.18$M_\odot$), in agreement with the  kinematic mass from the literature. As a low mass WD ($<$ 0.4$M_\odot$) necessitates formation through mass transfer (MT) in close binaries, WOCS1007 with a known period of 4.2 days along with its fast rotation, is likely to be formed by a case A or case B binary evolution. 

\end {abstract}

\section{Introduction}

In a star cluster, blue straggler stars (BSSs) are found to be brighter and bluer than stars that are close to the end of their main-sequence (MS) lifetimes, suggesting that these stars have continued to stay on the MS, defying further evolution. BSSs are believed to have gained mass resulting in a rejuvenation, though the process is not well understood. The dominant BSS formation mechanisms operating in both globular and open clusters are likely to be in some way dependent on binary stars (\citealp{Knigge2009}; \citealp{Mathieu2009Natur}, \citealp{Leigh2011}). Three processes are suggested in the literature: (1) stellar collisions in dynamical encounters between single, binary, triple systems (\citealp{Hills1976}, \citealp{Geller2013}) (2) transfer of material through Roche Lobe from a close companion in a binary (\citealp{McCrea1964}, \citealp{Tian2006}) and (3) a triple system where the doublet becomes a close binary, and merges to form a massive star (\citealp{Perets2009}, \citealp{Naoz2014}). The formation pathway by mass transfer (MT) in a binary produces a BSS with a initially hot companion, such as a white dwarf (WD). The direct observational evidence was obtained by \cite{gosnell2015}, by detecting WD companions to 7 BSSs in the old open cluster, NGC 188, and by \cite{2016Subramaniam} by detecting a post-AGB/HB companion to a BSS in the same cluster. In the case of globular clusters, \cite{Knigge2008} discovered the first BSS+WD binary in the central region of 47 Tuc and recently, \cite{Sahu2019} detected a BSS+WD system in the low density outer region of the globular cluster, NGC 5466.

M67 is very rich in BSSs, where 14 are confirmed as bona-fide members by \cite{Geller2015}. There have been several attempts previously, to detect hot companions to these BSSs, particularly from spectroscopic study using the \textit{International Ultraviolet Explorer} (\textit{IUE}) and UV photometry from the \textit{Ultraviolet Imaging Telescope}(\textit{UIT}). The seminal study by \cite{landsman98} used UIT images to detect 11 BSSs in M67. As there is a 0.5 mag uncertainty in the predicted UIT flux and 0.14 mag uncertainty in the measured UIT flux, they found strong evidence for a UV excess only in two BSSs (S975 (WOCS\footnote{WIYN open cluster study} 3010) and S1082 (WOCS2009, a triple system)). 
In particular, the absence of detection of a hot companion to WOCS1007 was surprising as this is a spectroscopic binary with a 4.2 day period and suspected to be currently undergoing MT \citep{milone1992f190}. The estimated companion mass of WOCS1007 was found to be 0.19 M$_\odot$ (M$_{2}$ $sin i$) by \cite{milone1992f190}, and was suggested as a BSS+WD by \cite{Shetrone2000}. It should be noted that low-mass WDs, with masses $<$0.4 M$_\odot$, are formed from stars that never ignited helium in their cores. As single star evolution takes more time than the age of the universe to form these WDs, formation of low mass WDs require evolution in close binaries \citep{Brown2010}. Hence detection and characterisation of the hot companion to WOCS1007 is extremely important to throw light on the formation pathways of both the BSS as well as the low mass WD.

The \textit{Ultra-violet Imaging Telescope} ({\it UVIT}) on board the Indian space observatory, {\it ASTROSAT}, has been producing far-UV (FUV) and near-UV (NUV) images of superior resolution of $\sim$1.5 arcsec, which is better than GALEX mission with its resolution of $>$4 arcsec in the same wavelength bands. Here we present analyses and results for 3 BSSs (WOCS1006, 1007 and 2011) in M67 observed in 3 FUV filters of {\it UVIT}. We combine the {\it UVIT} magnitudes with other estimations in the UV, optical and near-IR from space as well as ground observations to create multi-wavelength spectral energy distribution (SED). We present the details of data and observation in section 2, SED fits in section 3, and discussion in section 4.

\section{Observations and Data}

\begin{figure}
\centering
	\includegraphics[width=0.8\columnwidth]{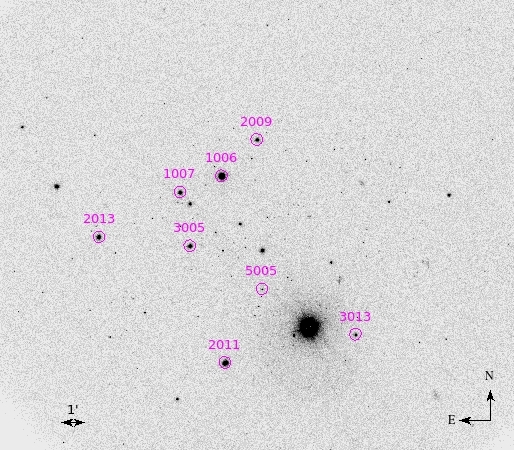}\\
\hspace{0.15cm}\includegraphics[width=0.25\columnwidth]{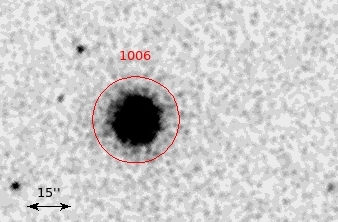}\hspace{0.25cm}\includegraphics[width=0.25\columnwidth]{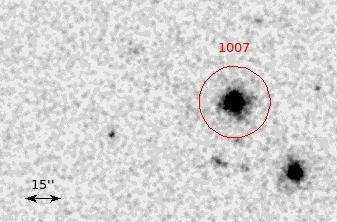}\hspace{0.25cm}\includegraphics[width=0.25\columnwidth]{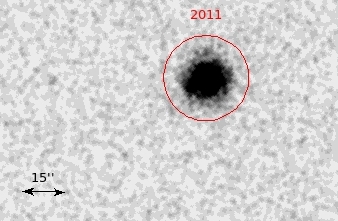}\hspace{0.15cm}\\
	\caption{ \textit{UVIT} image of M67 obtained in the F148W filter. A few stars are marked in the image and stamp size images of the three stars studied here are shown below.}
    \label{Fig: Two filter}
\end{figure}

M67 was observed in F148W, F154W and F169M filters, all observed on the same day. We were unable to get the NUV data due to payload related issues and the VIS data is useful only for drift correction. {\it UVIT} data were corrected for distortion, flat fielding and spacecraft drift using the {\sc CCDLAB} \citep{Postma2017}. %The science ready images were used for further analysis. 
Figure \ref{Fig: Two filter} shows the image in the F148W filter. Details of {\it UVIT}, in-orbit performance and calibration are described in \cite{Subramaniam2016Calb}, \cite{Tandoninorbit} and \cite{TandonCalb2017}.

\begin{table*}
	\centering
	 
	\caption{Details of all observations used in this study. The first column provides the date of observation (based on the availability). The filter details and exposure time are given in the second and third columns. 4th and 5th columns list the effective wavelength and bandwidth of the filter. The zero point magnitudes are listed in column 6.}
 
	\begin{tabular}{cccccc}
    \toprule
        Date of Observation & Filter & Exposure time (s) &$\lambda_{eff}$ (\AA) & $\Delta\lambda$	(\AA) &  m$_{zp}$\\
		\hline
		 2017 April 23 & {\it UVIT} F148W & 2290 & 1481 & 500  & 18.016\\
		 2017 April 23 & {\it UVIT} F154W & 2428 & 1541 & 380 &17.778\\
		 2017 April 23 & {\it UVIT} F169M &  2428 & 1608 & 290 & 17.455\\
         2006 Jan 21 & {\it GALEX} FUV &5555.2&1542&255&18.82\\
         2006 Jan 21 & {\it GALEX} NUV & 5555.2&2274&729&20.08\\
         
1986 March 18  & {\it IUE} NUV  &1200& 1900-3200 & 7* &\\
1986 March 19  & {\it IUE} NUV  &1800& 1900-3200 & 7* &\\
1986 March 20  & {\it IUE} NUV  &2100& 1900-3200 & 7* &\\

1986 March 18  & {\it IUE} FUV  &4500&1150-1950   & 7* &\\
1986 March 19  & {\it IUE} FUV  &7267&1150-1950   & 7* &\\
1987 Dec 15    & {\it IUE} FUV  &3600&1150-1950   & 7* &\\

         1990 Feb 15-18 & KPNO U  & 900&3630&592&\\
         1990 Feb 15-18 & KPNO B  & 480&4358&1004&\\
         1990 Feb 15-18 & KPNO V  & 240&5366&939&\\
         1990 Feb 15-18 & KPNO I  & 240&8100&1825&\\
           &{\it GAIA2} Gbp & & 5050&2347& 25.3806\\
         & {\it GAIA2}  G&&6230&4183&25.7934\\
         & {\it GAIA2} Grp&&7730&2756&25.1161\\
         1997 Nov 16 & 2MASS J & 403 &12350&1624&21.039\\
         1997 Nov 16 & 2MASS H & 403 &16620&2509&20.696\\
         1997 Nov 16 & 2MASS K & 403 &21590&2618&20.05\\
         2010 April 29 & WISE W1 & &33526&6626& 20.5\\
         2010 April 29 & WISE W2 & &46028&10422& 19.5\\
         2010 April 28 & WISE W3 & &115608&55055& 18.0\\     
 \hline
 
\multicolumn{6}{r}{* Spectral Resolution}
	\end{tabular}
\label{Tab:observation log} 
\end{table*}

PSF photometry was performed using standard {\sc IRAF} routines to obtain the magnitudes in all the filters, which are also corrected for aperture and saturation \citep{TandonCalb2017}. The limiting magnitude in all {\it UVIT} filters is $\sim$ 22 mag, with a maximum error of 0.2 mag. The {\it UVIT} observation details are listed in Table \ref{Tab:observation log}, along with already available observations, which are used in this study. As two are binaries, which can show photometric variability on timescales comparable to or shorter/longer than the exposure or cadence times, the date of observations are provided in table \ref{Tab:observation log}.

 \section{Spectral energy distribution}
The aim of this study is to check for the presence of hot companions to BSSs using the FUV flux detected by {\it UVIT}. Here we restrict our study to 3 of the confirmed 14 BSSs by \cite{Geller2015}. We study the potential candidate WOCS1007 for which a WD companion is expected based on its orbital solutions, along with WOCS1006 and WOCS2011, as these have IUE spectra. WOCS1006 and WOCS1007 are known to be single lined spectroscopic binaries and \cite{Mathys1991} have identified WOCS2011 as an Am star and \cite{Geller2015} through their radial velocity membership study have classified this star as a single member. WOCS1007 is a $\delta$ Scuti variable that shows pulsations in 26 frequencies with a maximum amplitude of 3 mmag at 230$\mu$Hz \citep{Bruntt2007}. \cite{Sindhu2018} inspected the IUE spectra and found the 2800 \AA\ MgII spectral line to be in absorption in all three stars, suggesting no/insignificant chromospheric activity in these stars. We check whether the detected FUV fluxes are in agreement with those expected for the BSSs, which in turn requires an accurate estimation of BSSs' properties. \cite{Sindhu2018} used SED analysis to estimate their properties, which are listed in their table 4. 

The {\it UVIT} fluxes (F148W, F154W \& F169M) are combined with fluxes from {\it GALEX} (\citealp{Martin05}; FUV \& NUV), {\it IUE} (\citealp{Boggess1978IUE01}; 1250 \AA, 1450 \AA, 1675 \AA, 2150 \AA, 2395 \AA, 2900 \AA\ - photometry from spectra), Optical (\citealp{montgomery93}; UBVRI), $Gaia$ (\citealp{Gaia2018}; G$_{bp}$, G, G$_{rp}$), 2MASS (\citealp{Cohen2003Calb2mass}; JHKs) and WISE (\citealp{Wright2010wise}; W1,W2 \& W3). Multi-wavelength SEDs with photometric flux from UV to IR listed in table \ref{Tab:flux}, spanning a wavelength range of 0.14 -11.5 $\mu$m, covered by a maximum of 23 data points, are constructed, after correcting for extinction in the respective bands \citep{Fitzpatrick1999}. We adopted E(B$-$V) = 0.041$\pm$0.004 mag \citep{Taylor2007}, distance modulus V$-$M$_{v}$ = 9.6$\pm$0.04 mag and Solar metallicity. We used the virtual-observatory SED analyser ({\sc VOSA}) to fit the SEDs and the details can be found in \cite{Sindhu2018}.
\begin{table*}
	\centering
	\caption{The Observed photometric flux and their respective error of the 3 BSSs detected in the three FUV filters of {\it UVIT} are listed along with GALEX - FUV \& NUV flux taken from GR6/GR7 data release and corrected for saturation, the optical flux from \cite{montgomery93} and IUE, 2MASS, WISE and GAIA taken from their respective source catalogue through VO photometry.}
		\begin{tabular}{cccccccc}
			\hline
	 WOCS ID	&  \multicolumn{2}{c}{1006}&\multicolumn{2}{c}{1007}& \multicolumn{2}{c}{2011}\\
		    Filter & Observed Flux & Observed Error  & Observed Flux & Observed error & Observed Flux & Observed Error\\
		    &(erg/s/cm$^2$/\AA)&(erg/s/cm2/\AA) & (erg/s/cm$^2$/\AA) & (erg/s/cm$^2$/\AA) &(erg/s/cm$^2$/\AA)& (erg/s/cm$^2$/\AA) \\    
			\hline
Astrosat/UVIT.F148W	&4.97E-14&	1.83E-16&	7.61E-15&	1.26E-16&	2.29E-14&	3.59E-16\\
Astrosat/UVIT.F154W&	5.32E-14&	7.84E-16&	8.17E-15&	1.20E-16&	2.53E-14&	2.57E-16\\
GALEX/GALEX.FUV	&5.53E-14&	1.19E-16&	6.13E-15&	2.58E-16&	2.14E-14&	2.86E-16\\
Astrosat/UVIT.F169M&	5.65E-14&	8.85E-16&	9.31E-15&	1.63E-16&	2.81E-14&	1.81E-16\\
IUE/IUE.1675-1725&	1.01E-13&	5.34E-15&	2.08E-14&	7.67E-15&	4.73E-14&	2.91E-15\\
IUE/IUE.2150-2200&	1.02E-13&	2.78E-14&	5.61E-14&	2.85E-14&	7.16E-14&	1.89E-14\\
GALEX/GALEX.NUV	&9.50E-14&	2.87E-16&	5.97E-14&	3.27E-16&	7.07E-14&	2.37E-16\\
IUE/IUE.2395-2445&	7.87E-14&	9.57E-15&	4.14E-14&	1.04E-14&	5.33E-14&	6.54E-15\\
IUE/IUE.2900-3000&	1.07E-13&	4.31E-15&	8.42E-14&	5.85E-15&	7.71E-14&	3.15E-15\\
KPNO/Mosaic.B&	2.27E-13&	6.77E-14&	2.15E-13&	1.20E-13&	1.70E-13&	4.96E-14\\
GAIA/GAIA2.Gbp&	1.69E-13&	5.05E-14&	1.64E-13&	9.17E-14&	1.25E-13&	3.64E-14\\
KPNO/Mosaic.V&	1.52E-13&	4.53E-14&	1.59E-13&	8.91E-14&	1.17E-13&	3.39E-14\\
GAIA/GAIA2.G&	1.08E-13&	3.22E-14&	1.13E-13&	6.33E-14&	8.02E-14&	2.33E-14\\
GAIA/GAIA2.Grp&	6.01E-14&	1.79E-14&	7.14E-14&	4.00E-14&	4.54E-14&	1.32E-14\\
KPNO/Mosaic.I&	4.93E-14&	1.47E-14&	6.12E-14&	3.43E-14&	3.95E-14&	1.15E-14\\
2MASS/2MASS.J&	1.57E-14&	2.61E-16&	2.10E-14&	3.50E-16&	1.22E-14&	1.92E-16\\
2MASS/2MASS.H&	5.65E-15&	9.38E-17&	8.29E-15&	1.53E-16&	4.38E-15&	7.26E-17\\
2MASS/2MASS.Ks&	2.24E-15&	3.72E-17&	3.34E-15&	4.93E-17&	1.72E-15&	2.53E-17\\
WISE/WISE.W1&	4.69E-16&	9.50E-18&	6.75E-16&	1.43E-17&	3.39E-16&	6.56E-18\\
WISE/WISE.W2&	1.32E-16&	2.55E-18&	1.95E-16&	3.59E-18&	9.76E-17&	1.88E-18\\
WISE/WISE.W3&	3.50E-18&	3.21E-19&	5.20E-18&	3.50E-19&		&\\

			\hline
	\end{tabular}
\label{Tab:flux} 
\end{table*}
The SED fits for the three stars are performed using the following three steps:

Step1: We first performed a single spectral fit to the SEDs using the Kurucz models (\citealp{Castelli1997} and updates) for the entire wavelength region, using the T$_{eff}$ and log $g$ as found by \cite{Sindhu2018}. The fluxes predicted by the model spectrum for the selected temperature and log $g$ for all the filters were estimated, which were then scaled to match the observed fluxes. $\chi^{2}_{red}$ values estimated for these single spectral fits are tabulated in the column 7 (first row of Table \ref{Tab:parameters}). We estimated the $\chi^{2}_{red}$ $^\dagger$ value {\footnote{ $\chi^{2}_{red}$ values calculated by ignoring filters with $\lambda_{eff}<1800$ \AA\ for a full single spectral fit (FUV to IR) are represented by "$\dagger$"}} for the same fit to compare with step2. This value is shown in the parentheses (first row).

Step2: As the $\chi^{2}_{red}$ values are large, we repeat step1, by fitting the SEDs excluding the photometric band pass shorter than 1800 \AA, to ignore any unusual UV flux and to refine the fit in the NUV-optical-IR region. The best fit $\chi^{2}_{red}$ $^\mathsection$ values {\footnote{$\chi^{2}_{red}$ values  obtained after a single spectral fit (NUV to IR) excluding filters with $\lambda_{eff}<1800$ \AA\ are represented by "$\mathsection$"}}  are tabulated in column 7 and second row of Table \ref{Tab:parameters}. The step2 fit is used to check for mismatch between the observed and the model SED in the shorter wavelengths and detect excess/deficient flux in the UV with respect to the expected BSS UV flux.

Step3: For both of the above fits, 
we estimated error weighted residual (EWR) flux using the following equation. 
\begin{equation}
\small
   EWR=(Flux_{Obs}-Flux_{Model})/Err_{Obs}
\label{residual}
\end{equation}
The SEDs of the BSSs are shown in figure \ref{Fig: SED of 9BSS}, and we discuss them below.

\subsection{WOCS1006} 
Step1 is performed using 23 photometric points with 2 fitting parameters and the number of degrees of freedom (Ndof) = 21. The SED fit shows a large $\chi^{2}_{red}$ value of 40.2 and a $\chi^{2}_{red}$ $^\dagger$ value of 11.2 (Fig \ref{Fig: SED of 9BSS}(a); EWR in the bottom panel). We repeat the fit using step2 (Fig \ref{Fig: SED of 9BSS}(b)) to obtain a better ${\chi^{2}_{red}}$ $^\mathsection$ value of 2.2. Step2 improves the flux residual in the optical and IR bands, with a slight excess in the UV region, as seen in one {\it UVIT} filter and GALEX FUV (Fig \ref{Fig: SED of 9BSS}(b) - bottom panel). As excess UV flux is not detected in the other two {\it UVIT} filters, we consider this star to be either a single BSS or a binary with a photometrically undetectable companion.

\subsection{WOCS2011} 
Step1 is performed using 22 photometric points with 2 fitting parameters (Ndof = 20). The SED fit shows a large $\chi^{2}_{red}$ (52.2) and $\chi^{2}_{red}$ $^\dagger$ values (21.5). The observed flux is more than the synthetic flux in the UV region, and vice versa in the longer wavelengths (Fig \ref{Fig: SED of 9BSS}(c)).  We repeat the fit using step2 (Fig.\ref{Fig: SED of 9BSS}(d)) and the best fit shows a lower $\chi^{2}_{red}$ $^\mathsection$ value of 0.9, suggesting a good fit and an improvement in the residual in the longer wavelengths. On the other hand, a deficiency in the FUV flux is found, as seen in the EWR plot (Fig \ref{Fig: SED of 9BSS}(d)). Step2 also increases the T$_{eff}$ from 8500 K to 8750 K. \cite{Nicolet1983} studied Am stars in the UV wavelength and observed significant deficiency of flux below 1800 \AA, which is in agreement with this study. Thus, WOCS2011 is likely to be a BSS with a deficient flux in FUV.

\subsection{WOCS1007}
\begin{figure*}
  \centering
  \begin{tabular}{c c}
    (a) & (b) \\
    \includegraphics[width=.43\linewidth, height=0.35\linewidth]{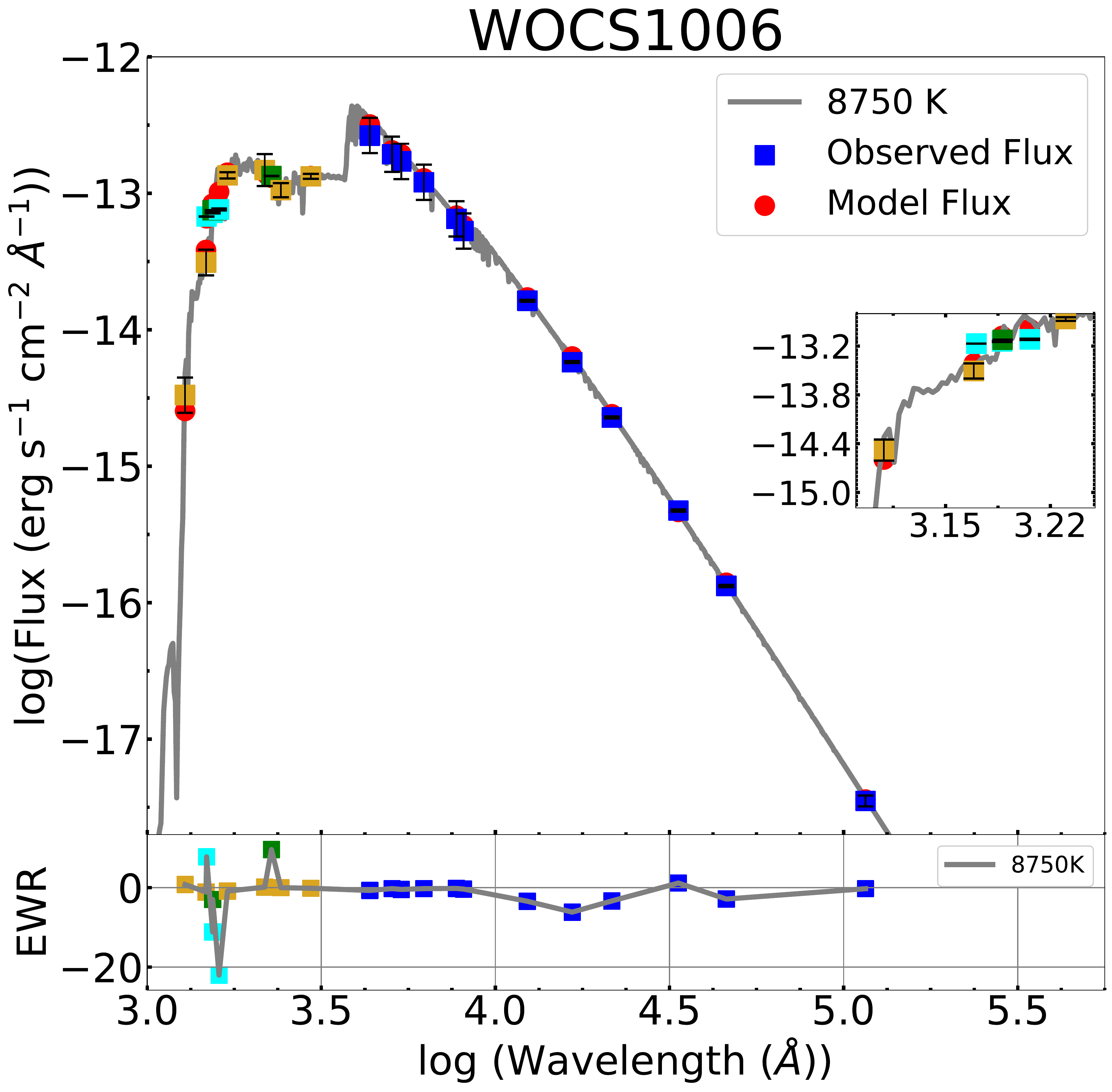} &
    \includegraphics[width=.43\linewidth, height=0.35\linewidth]{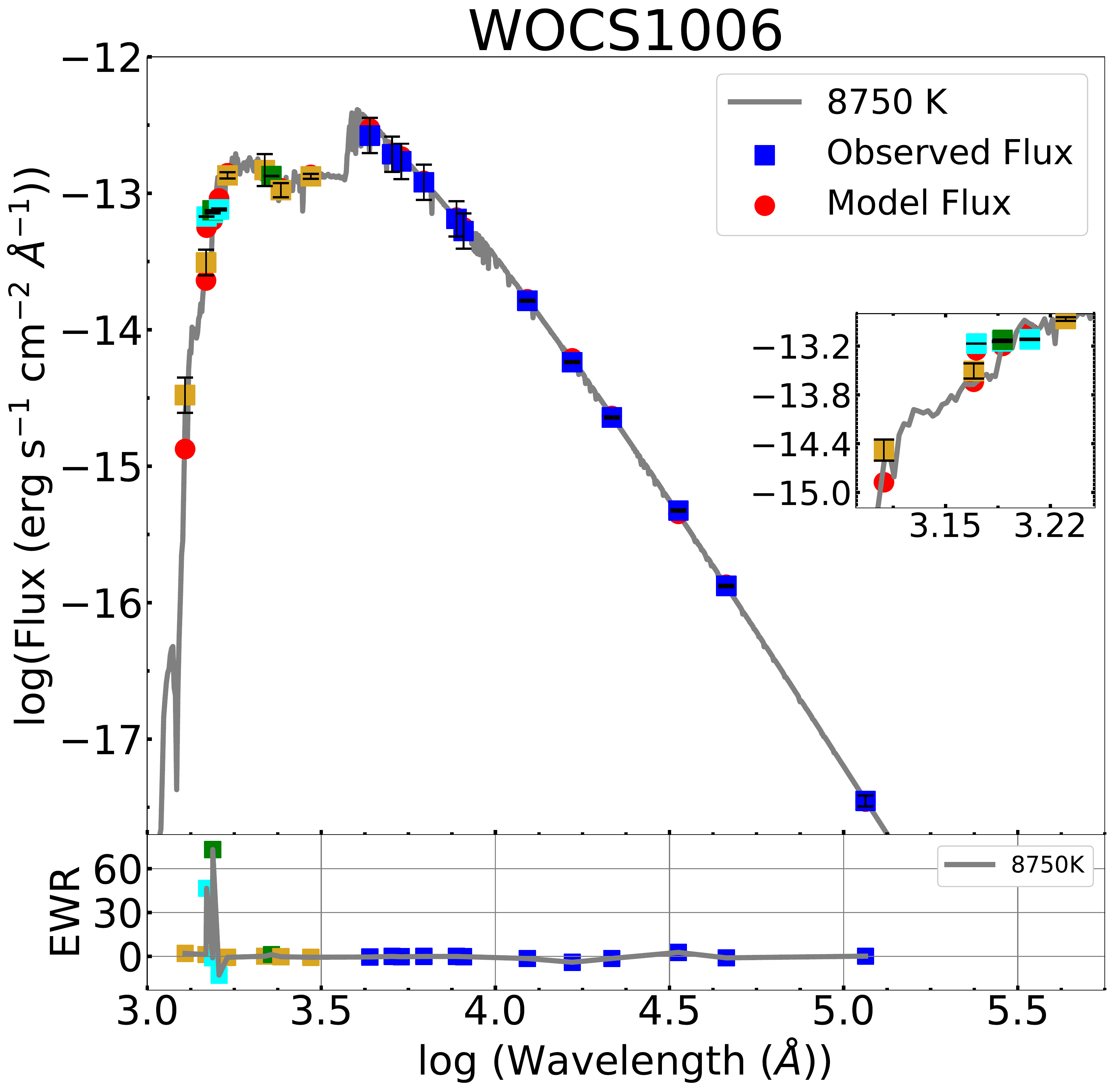} \\
    (c) & (d) \\
    \includegraphics[width=.43\linewidth, height=0.35\linewidth]{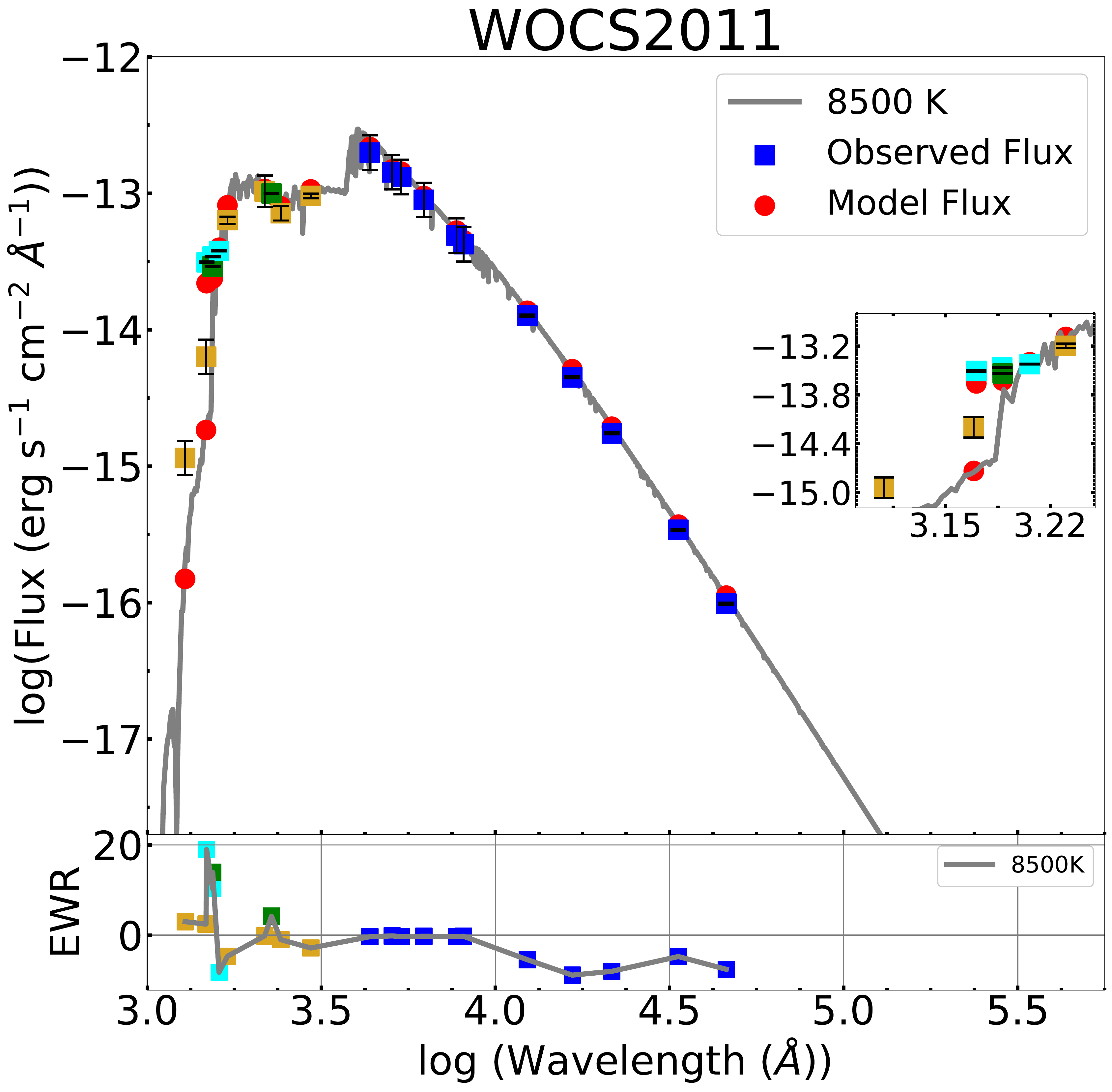} &
    \includegraphics[width=.43\linewidth, height=0.35\linewidth]{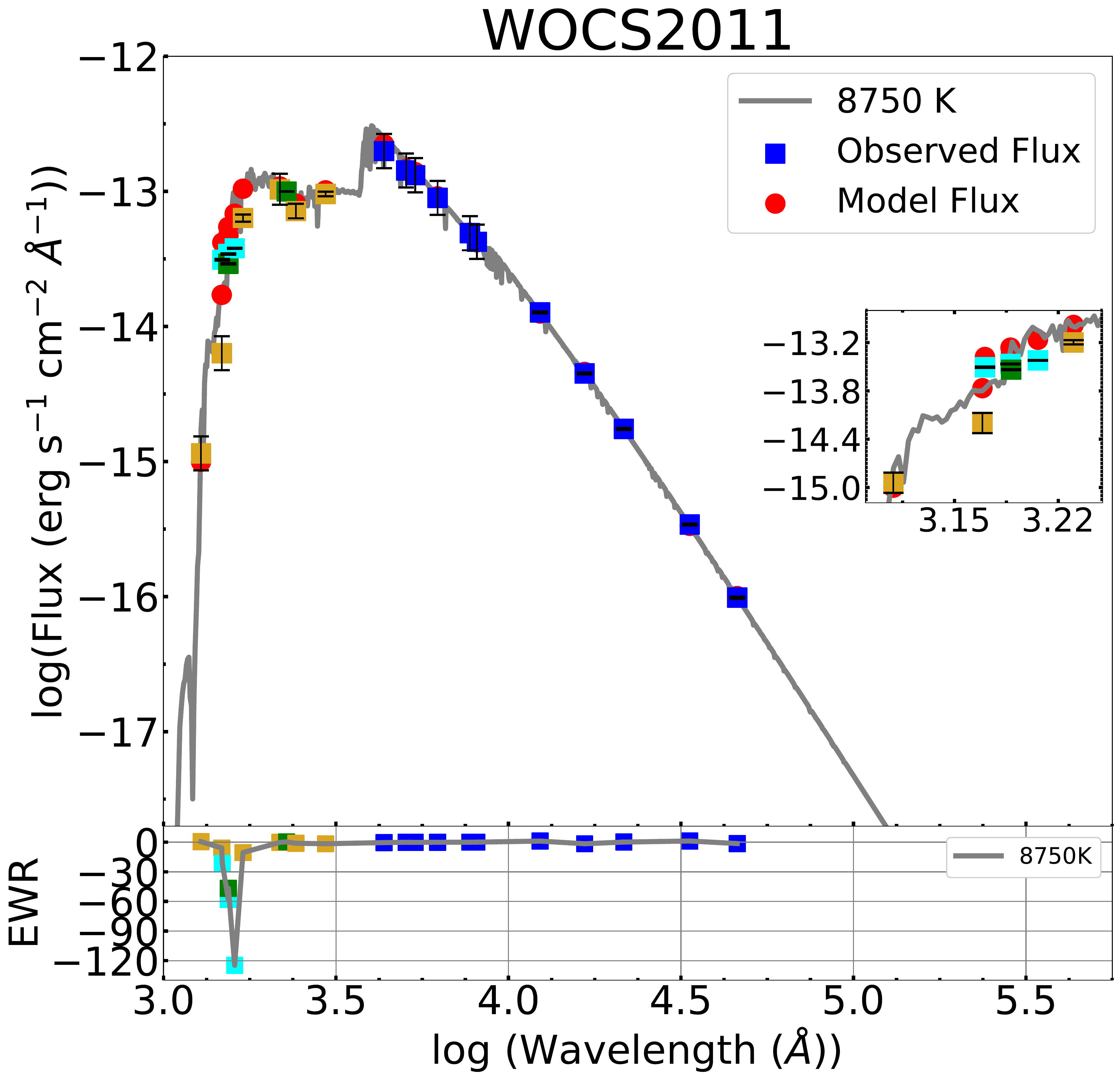} \\
    (e) & (f) \\ 
    \includegraphics[width=.43\linewidth, height=0.35\linewidth]{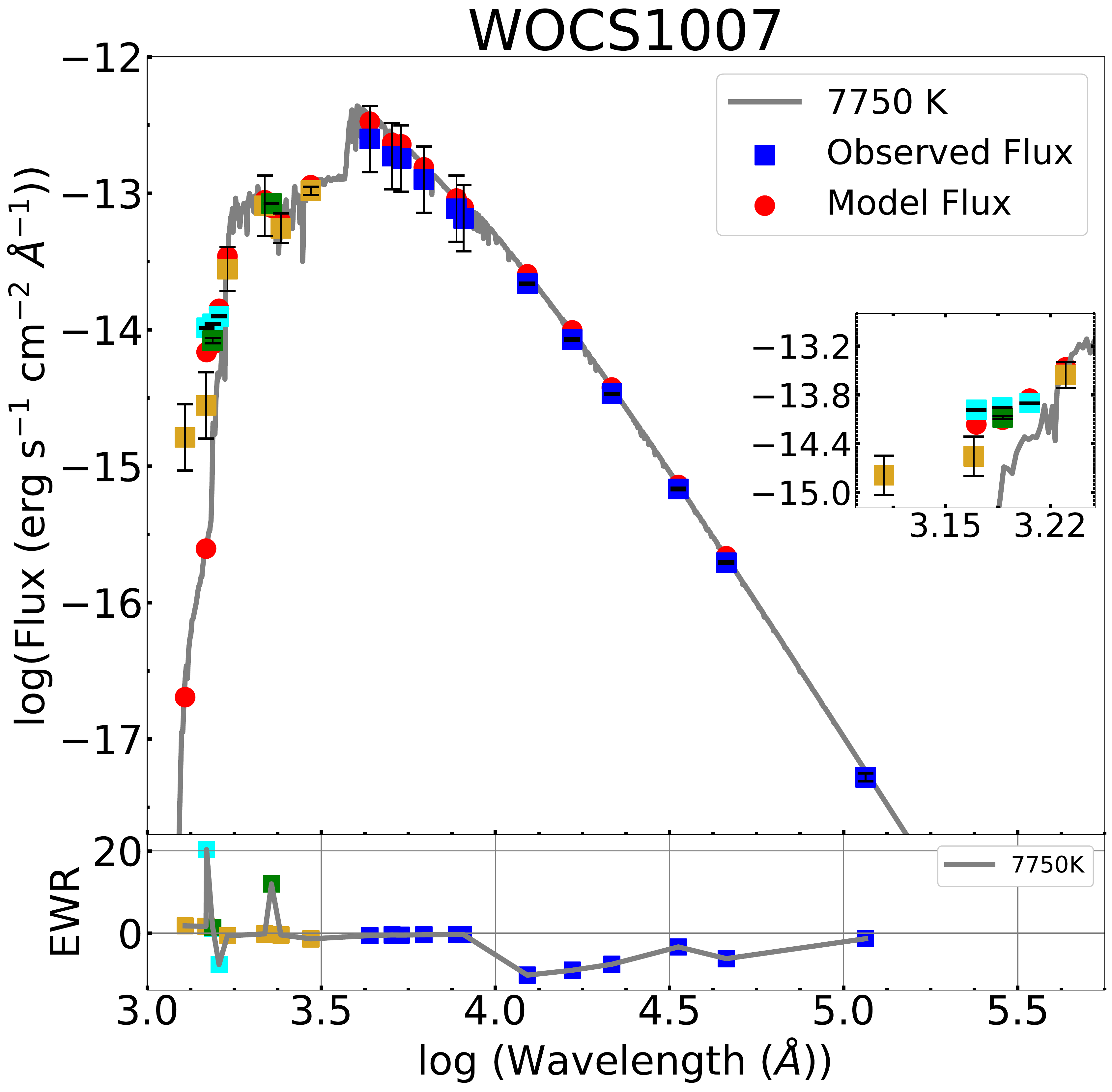} &
    \includegraphics[width=.43\linewidth, height=0.35\linewidth]{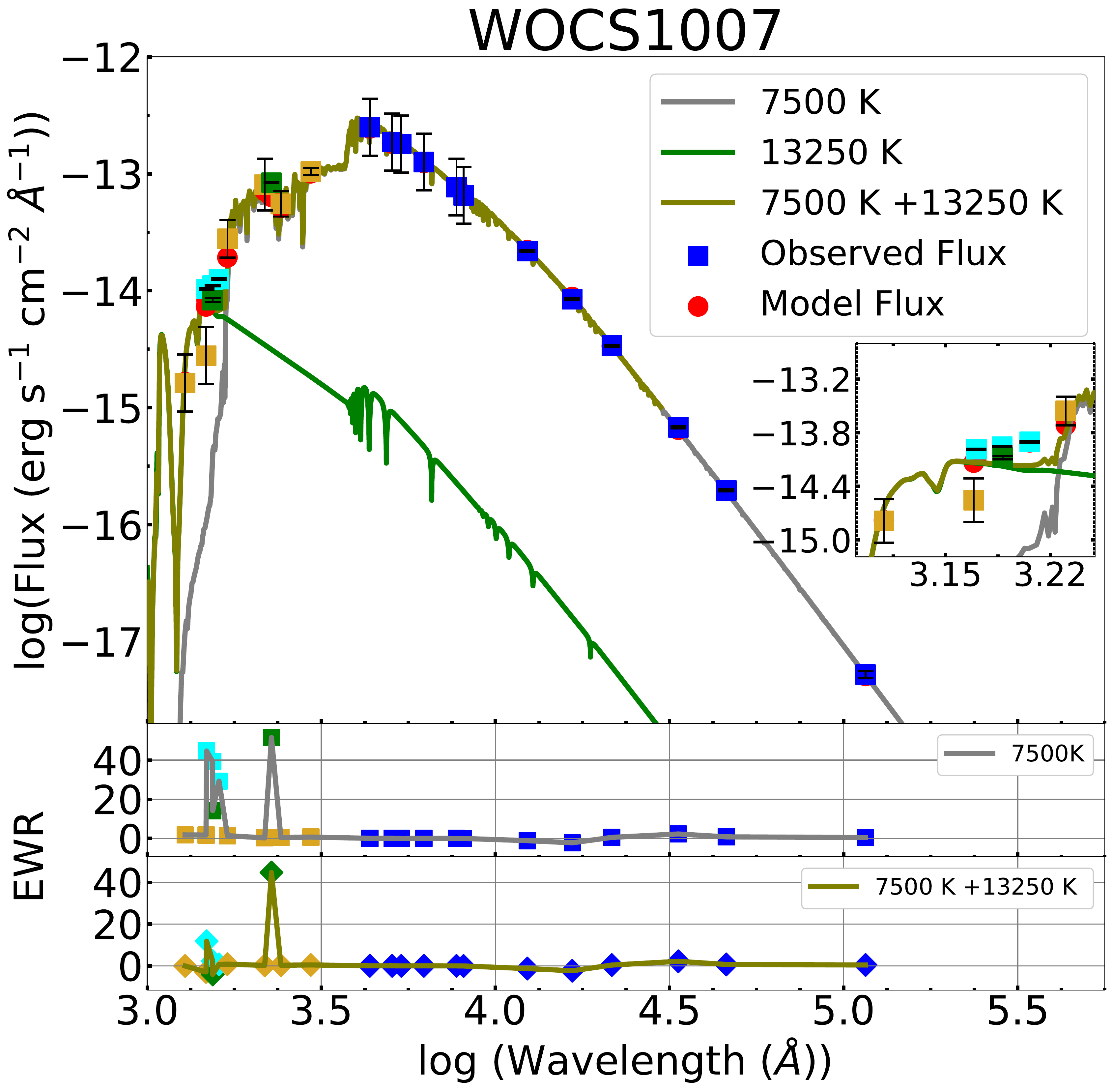} \\
  \end{tabular}
    \caption{SEDs of 3 BSSs with the close-up of the UV region in the insets. Left panels show step1 SED fits. Right panels show step2 SEDs for WOCS1006 and WOCS2011, and double fit for WOCS1007. Scaled and best fitting Kurucz spectrum (gray), Koester WD spectrum (green) and composite spectrum (olive) are shown along with corresponding T$_{eff}$. The observed photometric flux corrected for extinction are shown with blue squares (from optical to IR), Cyan ({\it UVIT}), green ({\it GALEX}) and golden ({\it IUE}) and corresponding composite synthetic flux as red dots. The EWR plots are in the bottom panels.  
 	} 
  \label{Fig: SED of 9BSS}
\end{figure*}

Step1 is performed using 23 photometric points with 2 fitting parameters (Ndof = 21). In the SED, the presence of an excess UV flux can be inferred from the observed data points extending significantly to the FUV region, when compared to the continuum shown in gray (Fig.\ref{Fig: SED of 9BSS}(e)). The SED fit shows a large $\chi^{2}_{red}$ (44.1) and $\chi^{2}_{red}$ $^\dagger$ (31.7) values. We repeat the fit using step2, which brings down the $\chi^{2}_{red}$ $^\mathsection$ (2.7). This fit results in the detection of a statistically significant excess flux  in all three {\it UVIT} FUV filters  and the {\it GALEX} FUV filter, ( Fig. \ref{Fig: SED of 9BSS}(f) - middle panel). As the residual (EWR) is weighted by error, the IUE data points do not show statistically significant excess due to large error, even though we detect $\sim$ 50\% more flux than those expected for the BSS in {\it UVIT, GALEX and IUE} filters. 
The residual plot of WOCS1007 also shows that the companion is likely to be a sub-luminous object (in comparison to figure 2 of \citealp{2016Subramaniam}).

We used Koester spectral models \citep{Koester2010, Tremblay2009} to fit the UV excess in the SED. In \cite{Kepler2015}, most of the WDs are found to have log $g \sim$ 8 (their Fig 4, 5 and 6). Therefore, we used three values of log $g$  (6.5, 7.75 and 9.0), though the broadband fluxes used in the SED are quite insensitive to the choice. As the fluxes of the hot component and BSS get added up, we created a composite SED in order to fit the full observed SED from UV to IR. 
The composite model SED is created by computing model fluxes for the BSS and WD, and adding the individually scaled fluxes.
Though the parameters (radius and temperature) of the WD spectra are mainly varied for the $\chi^{2}$ minimisation, minor modification of the BSS parameters were also done to obtain the overall good fit over the entire wavelength range. Adopting a range in extinction values from literature does not change the obtained parameters by more than 3 sigma.

We have used 21 data points for the double SED fit (Fig. \ref{Fig: SED of 9BSS}(f)) with 3 fitting parameters (Ndof = 18). F148W and \textit{GALEX} NUV points were not used, as the $\chi^2_{red}$ value increases due to small photometric errors, though the estimated parameters were not found to change significantly.  The EWR plot (bottom panel) suggests almost zero residual, except for one \textit{GALEX} NUV data point with large deviation due to small error. The reduction of $\chi^2_{red}$ value (44.1 to 2.5) favours the double fit, with $>$ 95\% confidence.

\begin{figure}
\centering
	\includegraphics[width=0.8\columnwidth]{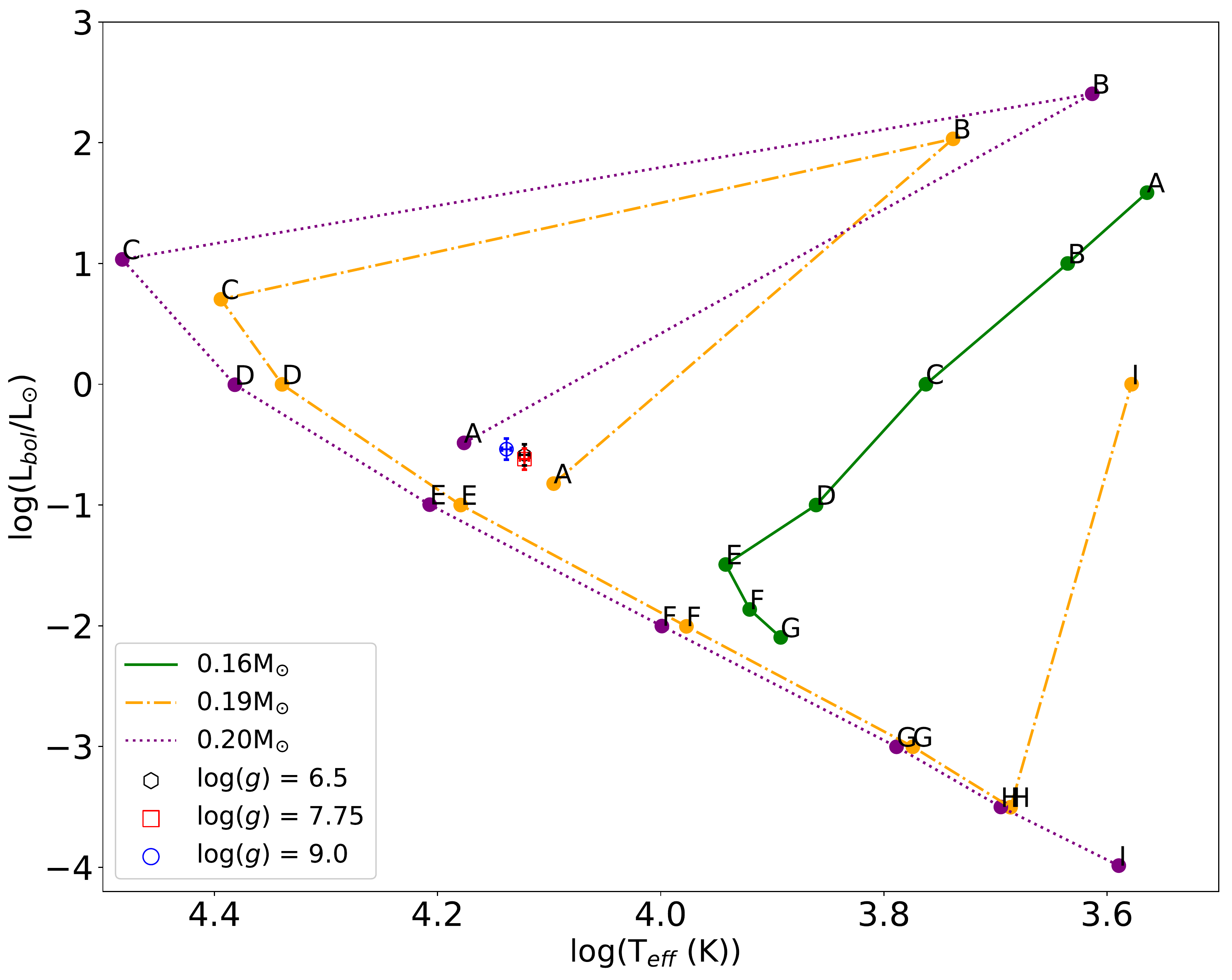}\\
	
	\caption{A H-R diagram of He-WD model taken from \cite{Panei2007} table 3 are plotted for 0.16M$_\odot$, 0.19M$_\odot$ and 0.20M$_\odot$. The characteristics of each labelled point are described in their table 3. The end of binary evolution (which is the starting point of WD evolution) is denoted by point A in the figure. The estimated parameters of WOCS1007 are shown in the plot for 3 log $g$ values 6.5, 7.75 and 9}
    \label{Fig:PaneiWD}
\end{figure}

We estimated the WD to have T$_{eff}$ = 13250 - 13750 K (depending on the log $g$), L/L$_\odot$ = 0.24-0.28 and R/R$_\odot$ = 0.094$\pm$0.01 -  0.097$\pm$0.01(assuming a 10\% error in distance to M67, as uncertainty of the distance with different distance modulus from various surveys). In fig. \ref{Fig:PaneiWD}, we have shown the evolutionary tracks for low mass helium WD models from \cite{Panei2007} for 0.16M$_\odot$, 0.19 M$_\odot$ and 0.20 M$_\odot$. 
The evolution of the WD begins after the mass loss phase at the point `A' labelled (which is also the end of binary evolution), in the fig. \ref{Fig:PaneiWD} for all three WD models. Further details of each point on the evolutionary track can be referred from their table 3. We have shown the star WOCS1007 with all three log $g$  values in the same figure. When we compare our estimated parameters to the WD models parameters, we find the closest match for low mass helium WD models, for a 0.19 M$_\odot$ or a 0.20 M$_\odot$ He WD. This suggests that the hot companion to WOCS1007 is likely to be a low mass WD.

 \begin{table*}
	\centering
	\caption{Fundamental parameters of the BSS and WD companion. The first and second columns list the WOCS and Sanders numbers, third, fourth, fifth and sixth columns list the T$_{eff}$, log $g$, Luminosity and Radius estimated for BSS (WOCS1006 and WOCS2011), and BSS \& WD companion (WOCS1007) respectively. $\chi^{2}_{red}$ for the single fit is listed in the 7th column. In the case of WOCS1007, the first row lists the $\chi^{2}_{red}$ for step1 fit, second row for step2 fit and the last three rows for the composite fit.}
	\vspace{0.2cm}
	\begin{tabular}{cccccccccc}
		\toprule %\hline
			WOCS ID & S No.& T$_{eff}$ (K) & log $g$& L/L$_\odot$&R/R$_\odot$ & $\chi^{2}_{red}$\\[4.5pt]
			\hline
			\multirow{2}{*}{1006} & \multirow{2}{*}{1066} & 8750$\pm$125&3.0 &26.75$\pm$0.72 & 2.31$\pm$0.03 & 40.2 (11.2$^\dagger$) \\
	& & 8750$\pm$125&3.5 &26.45$\pm$0.90 & 2.26$\pm$0.03 & 2.2$^\mathsection$ \\

\hline
			\multirow{2}{*}{2011} & \multirow{2}{*}{968}& 8500$\pm$125&4.0 &19.76$\pm$0.78 & 2.10$\pm$0.03 &52.2(21.5$^\dagger$)\\
			&& 8750$\pm$125&3.5 &19.79$\pm$0.54 & 1.96$\pm$0.03 & 0.9$^\mathsection$\\\hline
		\multirow{2}{*}{1007-BSS}&\multirow{2}{*}{1284}& 7750$\pm$125 &3.0 & 26.14$\pm$0.74& 2.97$\pm$0.04 &  44.1(31.7$^\dagger$) \\
		&  & 7500$\pm$125 & 3.5 & 24.57$\pm$1.01 & 2.94$\pm$0.04& 2.7$^\mathsection$\\\hline
		\multirow{3}{*}{1007-WD}& & 13250$\pm$125&  6.5&  0.26$\pm$0.05&0.097$\pm$0.01& 3.3\\
	                	&	&  13250$\pm$125& 7.75 &0.24$\pm$0.05   & 0.094$\pm$ 0.01	& 2.5\\ 
		              &	& 13750$\pm$125&9.0& 0.29$\pm$0.06&0.095$\pm$0.01 &2.4  \\
    \toprule
	\end{tabular}

	\label{Tab:parameters}
	\end{table*}

\section{Discussion}
\begin{figure*}
 		\centering
 	\includegraphics[width=130mm, scale=3.5]{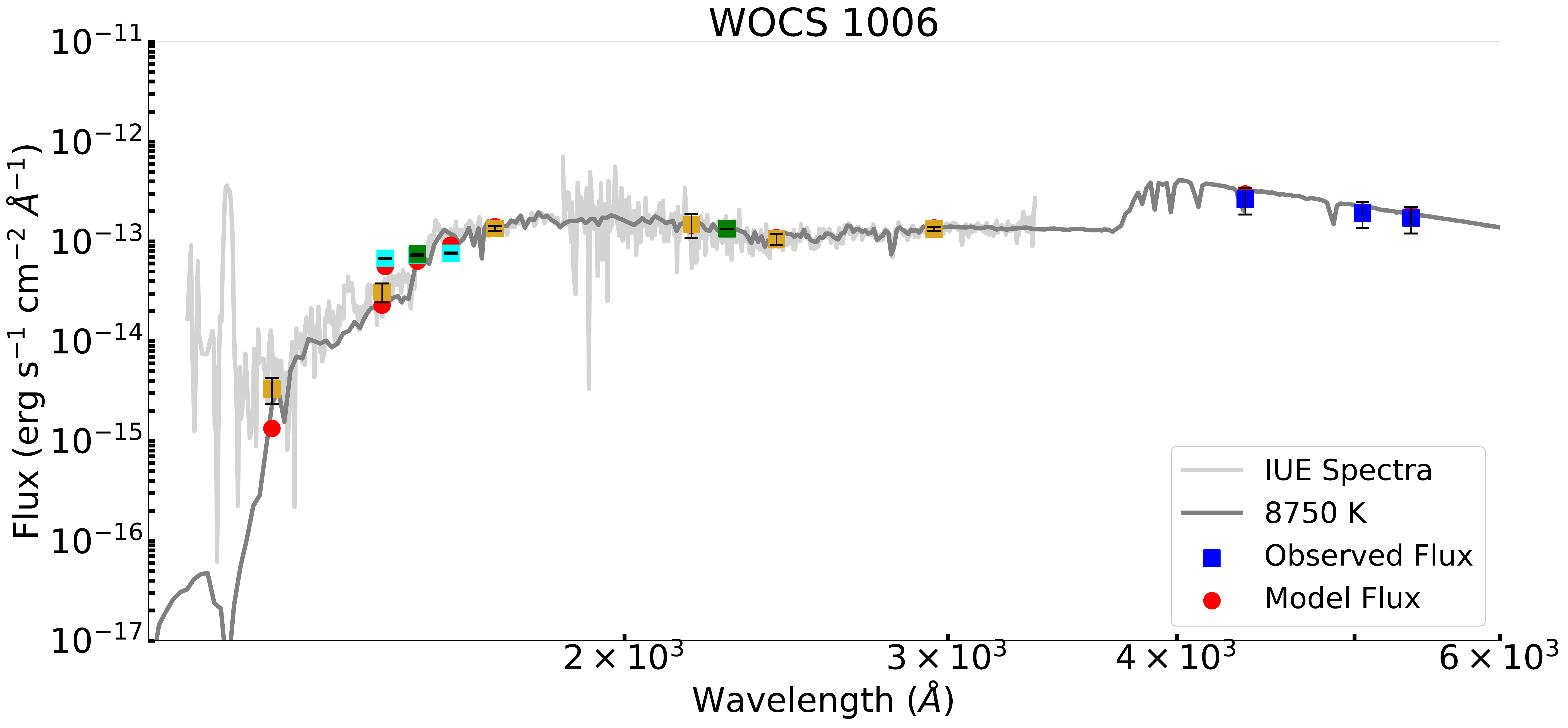} \\
	 \includegraphics[width=130mm, scale=3.5]{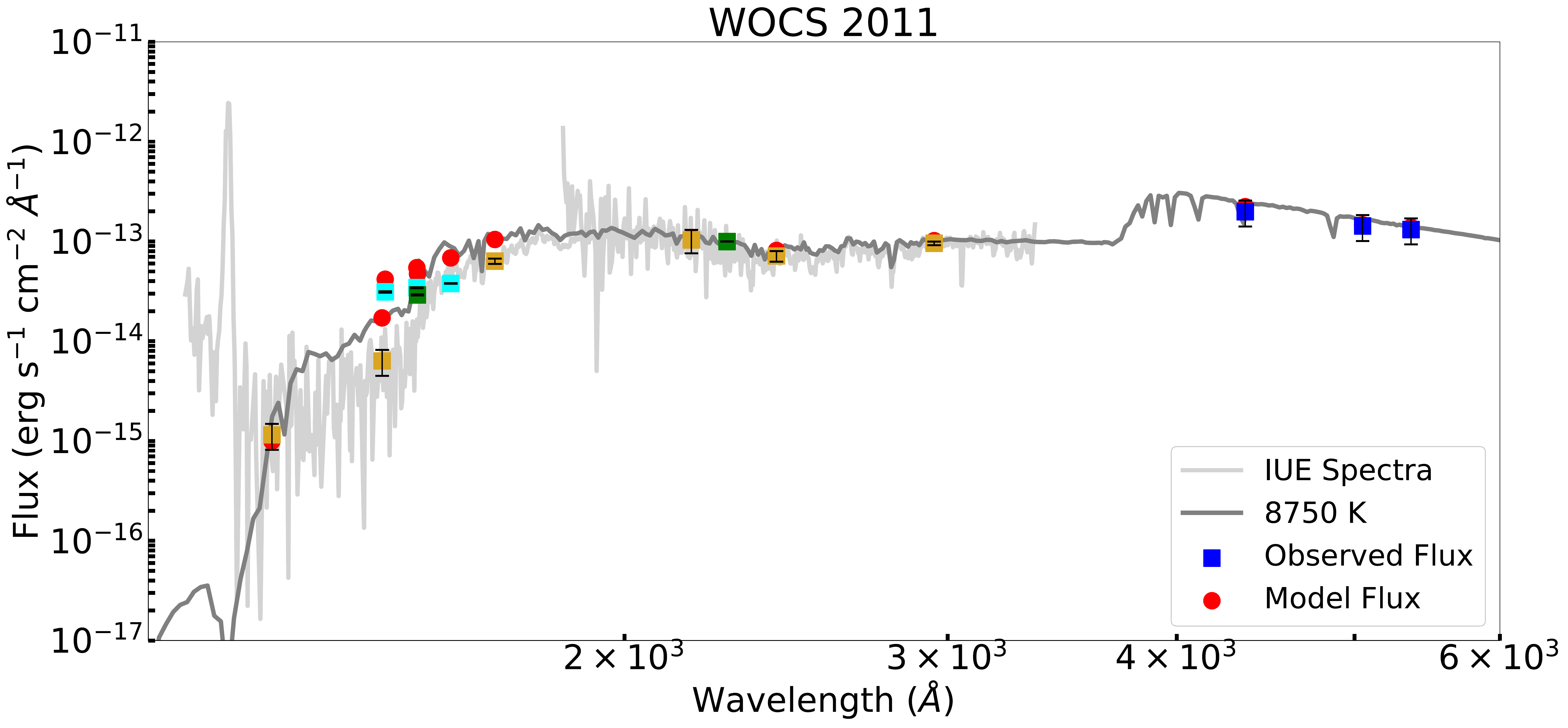}\\
 	  \includegraphics[width=130mm, scale=3.5]{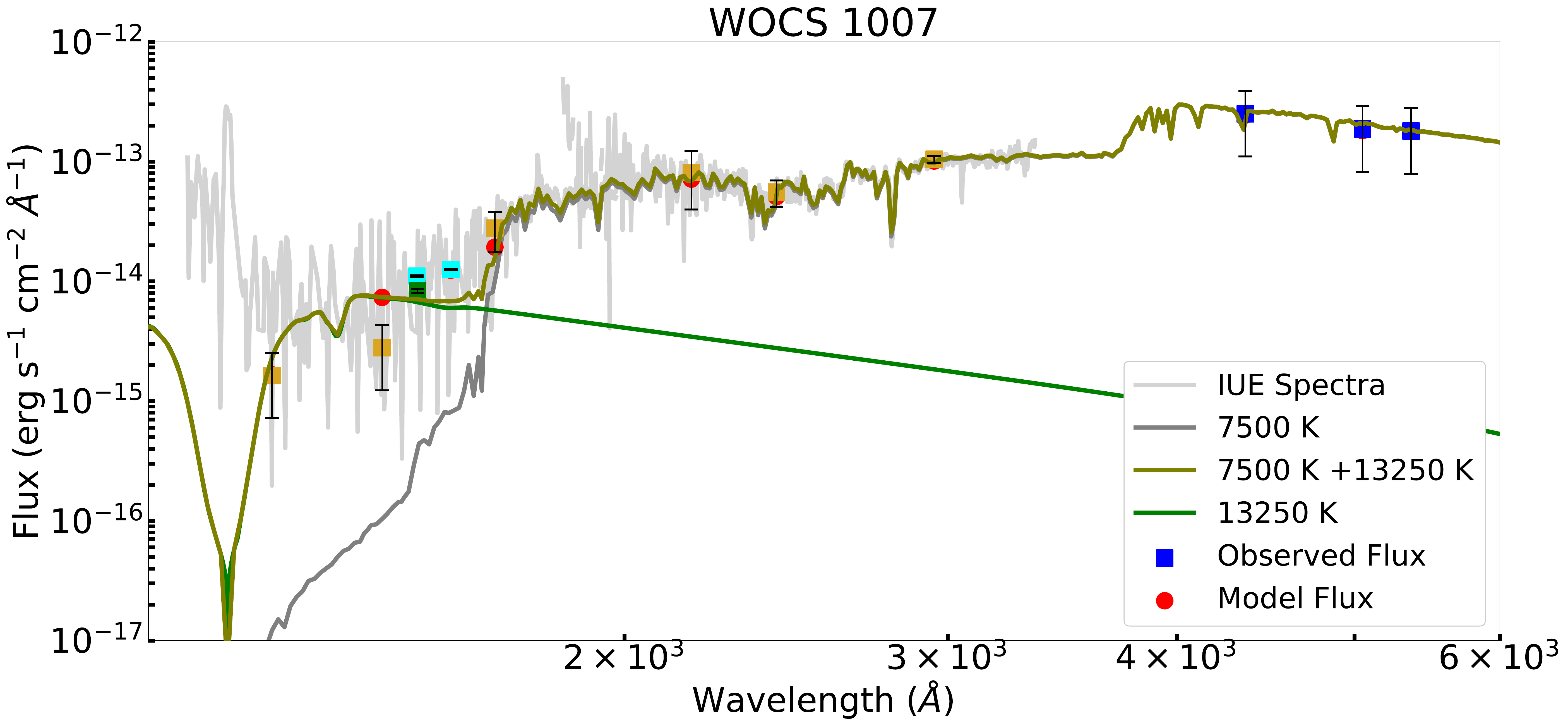}\\
 	\caption{SEDs of 3 BSSs (WOCS1006, WOCS2011 and WOCS1007) along with {\it IUE} spectra corrected for extinction. The colours codes are similar to fig. \ref{Fig: SED of 9BSS}.}
 	\label{Fig: SED of 3BSS with IUE}
 \end{figure*}

We have shown the observed and model estimated flux for {\it IUE}, {\it UVIT} and {\it GALEX} along with the observed extinction corrected spectra of {\it IUE} in figure \ref{Fig: SED of 3BSS with IUE}. The flux from the {\it UVIT} filters are consistent with the observed {\it IUE} spectra. The rising trend in the \textit{IUE} spectrum below 1250 \AA\ is due to geo-coronal Lyman alpha line, whereas UVIT filters do not detect it. 
Fig \ref{Fig: SED of 3BSS with IUE} confirms the presence of UV excess in the case of WOCS1007, as traced by the continuum of the {\it IUE} FUV spectrum. The {\it IUE} spectrum and the fitted spectral model for the hot component agree well, confirming the detection of the WD companion. In the case of WOCS2011, the {\it IUE} FUV spectrum also shows the presence of a deficiency in the FUV flux. 

WOCS1007 is a short period eccentric binary \citep{Latham1992} and a fast rotator ($v\ sini$ of 80 km s$^{-1}$ by \citealp{Milone1991etal}, 79.45 km s$^{-1}$ by \citealp{Bertelli2018}).
Our estimate of the secondary mass from the \cite{Panei2007} models suggests a $\sim$ 0.19 $-$ 0.20 M$_\odot$ WD. This is in agreement with the kinematic estimate of the secondary mass (M$_2$ $sin i$ = 0.19 - 0.21 M$_\odot$) by \cite{Latham1992}, for a primary (M$_1$) mass of 2.0 - 2.2 M$_\odot$. This makes the secondary companion belong to the class of low mass WDs, which are formed only in close binaries, as single star evolution prohibits the formation of WDs with mass less than 0.4 M$_\odot$ within the Hubble time \citep{Brown2010, Istrate2016}. The accepted mechanism for the formation of low-mass He-core WDs is either through unstable mass loss via common-envelope episodes or stable mass loss via Roche-lobe overflow in close binaries \citep{Istrate2016, Calcaferro2018}. 

The WD companion of WOCS1007 is consistent with He WD models, with a luminosity of 0.1 - 0.15L$_\odot$ which is within 3$\sigma$ from our estimated value. In Fig. \ref{Fig:PaneiWD}, the WD lies close to the points labelled `A' and 'E' of the 0.19 and 0.20 M$_\odot$ models. If we consider the WD is at the point 'A', then the WD is recently formed and is still evolving towards a typical WD. It also demands the MT to have stopped very recently, as supported by the large rotation of the BSS. On the other hand, if we consider that the WD is at the evolutionary point 'E', then the WD is formed about $\sim$ 190 Myr ago for a 0.19 M$_\odot$ or $\sim$ 160 Myr for a 0.20 M$_\odot$.

\cite{Lu2010} studied the formation of BSSs via MT in close binaries in M67. They followed the evolution of a close binary ($\sim$ 1.4 days with circular orbit) of 1.4 M$_\odot$ + 0.9 M$_\odot$ and compared the evolutionary behaviour of case A and B MT cases. The end product of the close binary pair was found to be a 2.04 M$_\odot$ BSS with a 0.26 M$_\odot$ WD companion, which is very similar to the WOCS1007 system. Presently, the BSS is estimated to have a mass of 2.0 - 2.2 M$_\odot$, which demands that it has gained at least 0.7 - 0.9 M$_\odot$ (assuming its progenitor to be a 1.3M$_\odot$ star and more, if it is less massive). The progenitor of the WD, expected to be $\sim$1.35M$_\odot$, would have therefore transferred at least 0.7 - 0.9 M$_\odot$ to the secondary during the MT, which is $\ge$50\% to its original mass. This demands an efficient MT and hence a case A/B MT may be preferred to other types of MT, such as, wind accretion model \citep{Perets2015}. A detailed simulation of WOCS1007 is necessary to derive the details of the progenitors as well as the mode of MT which created the present configuration.

WOCS2011 is known to be an Am star \citep{Mathys1991} and is a slow rotator ($\sim$ a few km s$^{-1}$ by \citealp{Milone1991etal}), 
and could have slowed down due to magnetic breaking. We find the star to have deficient flux in FUV, which is typical for the Am stars. On the other hand, it is likely to have no/less chromospheric activity as indicated by the 2800\AA\ MgII absorption line. 
 
WOCS1006 is one of the fast rotating BSS with a $v\ sini$ of 100 km s$^{-1}$ \citep{Milone1991etal}. WOCS1006 does not have a detectable WD companion, though it is known to be a single lined spectroscopic binary. It is possible that this is also a post-MT object, as suggested by its fast rotation, where the WD companion has cooled enough by now and is no longer detectable, and there is no magnetic activity for the BSS to spin down. It should be noted that there is a limited window to actually be able to detect these WD companions photometrically before they cool down and become too faint relative to the BSSs. It is thus essential to monitor their radial velocities to detect companions and constrain their masses, wherever possible. 
 
This study presents the first detection of a low mass He WD as a companion to a BSS. Previous detections of such WDs in binary systems are mostly in double degenerate systems, where the companion is either a neutron star or a normal WD (\citep{Liebert_2004, Vennes_2011} and references therein). It is also the first time such a low mass WD is detected in the well studied M67 cluster. This discovery will therefore help identifying/constraining the formation pathways of not only the low mass He-WDs but also the BSSs. \\

{\bf Summary:} We report the first detection of a WD companion to one of the BSSs (WOCS1007) in M67, using FUV images from the {\it UVIT} on {\it ASTROSAT}. The WD companion is found to have a T$_{eff}$ $\sim$ 13250-13750K and a radius of 0.09 R$_\odot$, comparable to a He WD of 0.18 M$_\odot$, confirming it to be a low mass WD. As single star evolution cannot produce a low mass WD within the Hubble time, formation through MT in close binaries is necessary. Thus, we suggest that WOCS1007 is formed as a result of a MT in a close binary, possibly through a case A or case B binary evolution. We also detect a deficiency in the FUV flux for WOCS2011, which is an Am star.
\\
\\
\textit{UVIT} project is a result of collaboration between IIA, Bengaluru, IUCAA, Pune, TIFR, Mumbai, several centres of ISRO, and CSA.
This publication makes use of {\sc VOSA}, developed under the Spanish Virtual Observatory project supported from the Spanish MICINN through grant AyA2008-02156. SN acknowledges support from CSIR for the grant 09/890(0005)/17 EMR-I. We thank the anonymous referee for the useful comments.

\bibliography{references}

%\appendix
%\section{Appendix information}

% \subsection{Conclusion}
  
\end{document}